\documentclass[aps,prl,twocolumn,showpacs,preprintnumbers,amsmath,amssymb]{revtex4}
\usepackage{graphicx}% Include figure files
\usepackage{dcolumn}% Align table columns on decimal point
\usepackage{bm}% bold math

\begin{document}

%\preprint{APS/123-QED}

\title{The $^{87}$Sr optical frequency standard at PTB}% 
\author{St.~Falke, H.~Schnatz, J.S.R.~Vellore Winfred, Th.~Middelmann, St.~Vogt, S.~Weyers, B.~Lipphardt, G.~Grosche, F.~Riehle, U.~Sterr and Ch.~Lisdat}
\email{christian.lisdat@ptb.de}

\affiliation{Physikalisch-Technische Bundesanstalt, Bundesallee 100, 38116 Braunschweig, Germany}

\date{\today}% It is always \today, today,

\begin{abstract}
With $^{87}$Sr atoms confined in a one dimensional optical lattice, the frequency of the optical clock transition $5s^2$ $^1$S$_0$ -- $5s5p$ $^3$P$_0$ has been determined  to be 429\;228\;004\;229\;872.9(5)~Hz. The transition frequency was measured with the help of a fs-frequency comb against one of PTB's H-masers whose frequency was measured simultaneously by the PTB Cs fountain clock CSF1. The Sr optical frequency standard contributes with a fractional uncertainty of $1.5 \times 10^{-16}$ to the total uncertainty. The agreement of the measured transition frequency with previous measurements at other institutes supports the status of this transition as secondary representation of the second with the currently smallest uncertainty.
\end{abstract}

% PACS, the Physics and Astronomy Classification Scheme.

%\preprint{1.0}
\maketitle

%
%
%%%%%%%%%%%%%%%%%%%%%%%%%%%%%%%%%%%%%%%%%%%%%%%%%%%%%%%%%%%%%%%%%%%%%%%%%%%%%%%%%%%%%%%%%%%%%%%
%
%	Introduction
%
%%%%%%%%%%%%%%%%%%%%%%%%%%%%%%%%%%%%%%%%%%%%%%%%%%%%%%%%%%%%%%%%%%%%%%%%%%%%%%%%%%%%%%%%%%%%%%%
%
%
\section{Introduction} \label{sec:intro}

The development of frequency standards, both, in the microwave \cite{lev10, wey09, cha08a} and optical \cite{lud08, ros08, mar04, sch05a} domain is a very active field of research. The investigations reach out far beyond the aspects of metrology into questions of fundamental research \cite{swa11, cam09, cho10a, lor08, szy07}. The development of optical clocks using optical transitions as reference has provided a number of systems \cite{cip09, gil06b} that have a stability far better than today's Cs fountain clocks. Several ones have already the capability to achieve a systematic uncertainty better than the best realizations of the second \cite{lud08, ros08, sch05a}.

The field of optical clocks can be divided into frequency standards with single ions (or as proposed few ions) and with a large number ($\sim$$10^4$) of neutral atoms. Single ion standards lead the race for the highest accuracy \cite{ros08}. Standards with neutral atoms, however, promise an extraordinary stability due to the larger number of absorbers that are simultaneously interrogated. Among these, systems using the fermion $^{87}$Sr as reference are widely distributed and are investigated to our knowledge in at least eight laboratories. 

The newly developed optical clocks with high stability allow measuring frequency ratios without any reference to the realization of the second by Cs clocks. With stabilized optical links  \cite{lop10, pap10, hon09} these measurements can in principle be performed between distant clocks. However, they do not yet connect optical frequency standards  based on the same transition in different institutions to test their accuracy beyond the limit set by primary clocks. In this respect, only a few high accuracy tests of such standards have been reported so far \cite{sch05a,cho10}; and these have been within one lab. Frequency measurements of optical frequency standards against primary clocks are therefore of high interest to evaluate the performance of distant systems.

So far, for $^{87}$Sr lattice clocks high accuracy measurements have been performed in three laboratories \cite{cam08b, hon09, bai08}. All results show an excellent agreement, which has resulted in the acceptance of the $5s^2$ $^1$S$_0$ -- $5s5p$ $^3$P$_0$ clock transition as secondary representation of the second with currently the smallest uncertainty \cite{cip09}.

We present here the first frequency measurement of this transition performed with the Sr optical frequency standard at the Physikalisch-Technische Bundesanstalt (PTB). An uncertainty of 0.5~Hz or in fractional units of $1.0 \times 10^{-15}$ has been achieved, confirming the previously published results.

The paper is organized as follows: In Sec.~\ref{sec:setup} the experimental setup and interrogation sequence is outlined in detail. Systematic shifts of the clock transition and their uncertainty are discussed in Sec.~\ref{sec:shift} followed by the frequency measurement and its results in Sec.~\ref{sec:frequ}.
%
%
%%%%%%%%%%%%%%%%%%%%%%%%%%%%%%%%%%%%%%%%%%%%%%%%%%%%%%%%%%%%%%%%%%%%%%%%%%%%%%%%%%%%%%%%%%%%%%%
%
%	Experimental setup
%
%%%%%%%%%%%%%%%%%%%%%%%%%%%%%%%%%%%%%%%%%%%%%%%%%%%%%%%%%%%%%%%%%%%%%%%%%%%%%%%%%%%%%%%%%%%%%%%
%
%
\section{Experimental setup} \label{sec:setup}
\begin{figure}
\centerline{\includegraphics[width=1\columnwidth]{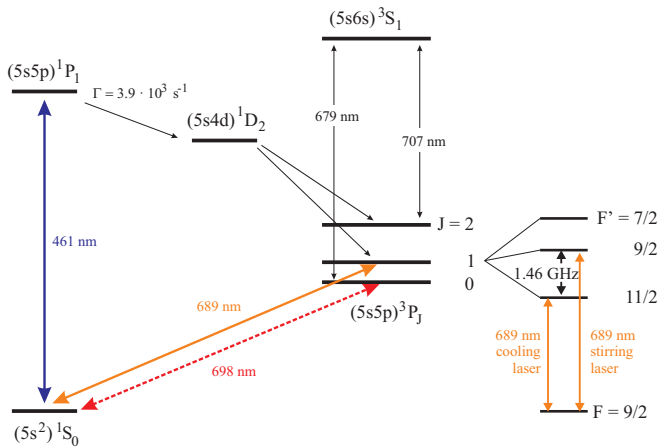}}
\caption{Partial level scheme of $^{87}$Sr with the two cooling transitions (461~nm and 689~nm), repumpers (679~nm and 707~nm), and the clock transition at 698~nm. The hyper-fine structure splitting of the $^3$P$_1$ level is indicated on the right.}
\label{fig:level}
\end{figure}

Parts of the experimental setup have been described before \cite{leg09, lis09, mid11}. For completeness, we give a detailed description of the frequency standard as used to perform the absolute frequency measurement: $^{87}$Sr atoms are evaporated from a furnace at a temperature of 520~$^\circ$C, collimated by a 2~mm diameter pinhole. They are laser cooled in a Zeeman slower and then deflected and collimated by a two dimensional optical molasses operated with light red detuned by 385~MHz (slower) and 10~MHz (molasses) from the 461~nm transition $^1$S$_0$ -- $^1$P$_1$ (for the relevant atomic levels see Fig.~\ref{fig:level}). The deflection of the cooled atomic beam into the capture region of the magneto-optical trap (MOT) ensures that no direct line-of-sight exists between the hot furnace and the trapping region.  Additionally, the room-temperature pinhole between furnace and Zeeman slower reduces the indirect blackbody radiation flux in the direction of the MOT.

The MOT with a magnetic field gradient of 7.4~mT/cm is operated with light detuned by 32~MHz below the 461~nm transition (natural line width: 32~MHz). The 1/${\rm e}^2$ beam radii of the MOT beams are 5~mm and they have peak intensities of about 3.5~mW/cm$^2$ each. In the MOT a cloud of $^{87}$Sr atoms is cooled within 300~ms to about 3~mK. All laser beams are derived from a frequency doubled diode laser system whose frequency is stabilized to a long-term stable reference resonator. To increase the atom number, repump laser beams resonant with the 707~nm $^3$P$_2$ -- $^3$S$_1$ and 679~nm $^3$P$_0$ -- $^3$S$_1$ transitions are applied. To cover the hyperfine manifold of the $^3$P$_2$ level the 707~nm repump laser is rapidly scanned over an interval of 5~GHz. Its average frequency is stabilized with a commercial wavemeter. The 679~nm repump laser is frequency stabilized to a reference resonator.

The pre-cooled atoms are then reloaded into a MOT with lowered magnetic field gradient of 0.7~mT/cm operated on the 689~nm intercombination line $^1$S$_0$ -- $^3$P$_1$ with a natural linewidth of 7~kHz. The laser is tuned 1.6~MHz below the resonance of the $\Delta F = F' - F = +1$ transition (with the total angular momentum  $F$) and is frequency modulated at 30~kHz with a peak-to-peak excursion of 3~MHz to increase the velocity capture range. The MOT beams have 1/${\rm e}^2$ beam radii of 2.6~mm and peak intensities of about 5.5~mW/cm$^2$. Beams from a second laser (the so-called stirring laser) are superimposed with the cooling laser beams. The beam intensity, detuning, and frequency modulation with respect to the 689~nm $\Delta F = 0$ transition are similar to the one of the cooling laser. Stirring light is applied to redistribute the population in the Zeeman sublevels and thus enhance the cooling efficiency, which is otherwise reduced by optical pumping to untrapped levels. This is due to strongly different $g$-factors of the $^1$S$_0$ and $^3$P$_1$ levels of the cooling transitions that cause large variations in the detuning of the cooling beams from the Zeeman transitions (and thus excitation probabilities) that are not compensated by the Clebsch-Gordon coefficients \cite{muk03}. After a 90~ms long cooling phase a temperature of about 50~$\mu$K is reached. The temperature is further lowered within 50~ms to $\sim$5~$\mu$K applying only the cooling laser beams without frequency modulation at reduced detuning (400~kHz) and intensity (440~$\mu$W/cm$^2$).

A horizontally oriented and linearly polarized 1D optical lattice is overlapped with the atomic cloud during the whole cooling sequence. The lattice is formed by a retro-reflected laser beam from a Ti:Sapphire laser at the magic wavelength at 813~nm \cite{kat03}. The frequency of the lattice laser is stabilized to a fs~frequency comb and monitored with a wavemeter with 1~MHz accuracy to unambiguously detect mode hops of the lattice laser followed by relocking to a different fs-comb mode. The lattice waist radius is about 32~$\mu$m; the trap depth is typically $U_T = 125\; E_r$, where $E_r = \hbar^2 k^2 / 2m$ is the photon recoil energy with the wavevector $k = 2 \pi /\lambda$ of the lattice light and $m$ is the atomic mass of $^{87}$Sr. Motional sidebands in the lattice are observed with frequencies of $\nu_z \approx 77$~kHz in axial and $\nu_r \approx 440$~Hz in radial direction. 

After turning off the light of the MOTs the atoms in the lattice have a temperature similar to the one in the last MOT phase. The atoms in the lattice are then spin-polarized to either the $m_F = +9/2$ or $m_F = -9/2$ $^1$S$_0$ Zeeman level by an appropriately circularly polarized light pulse of 1~ms length resonant with the $\Delta F = 0$ $^1$S$_0$ -- $^3$P$_1$ transition. To define the quantization axis, the MOT quadrupole coils are used in approximately Helmholtz configuration to generate a homogeneous magnetic field of about 23~$\mu$T. The magnetic field axis is parallel to the lattice polarization. After spin-polarization, the magnetic field is increased to 1.8~mT and an intense $\pi$ pulse of around 4.8~ms duration, resonant with the $^1$S$_0$ -- $^3$P$_0$ $m_F = +9/2$ or $m_F = -9/2$, $\Delta m_F = 0$ clock transition, is applied. The clock laser beam is aligned with the axis of the optical lattice with a residual angle of less then 0.2~mrad. It has a 1/${\rm e}^2$ radius of about 65~$\mu$m and has linear polarization that is parallel to the orientation of the magnetic field and the linear lattice polarization. 

The light is delivered via a polarization maintaining optical fibre from an extended cavity diode laser in Littmann configuration. The laser frequency is offset by an acousto optical modulator (AOM) and is stabilized to a resonator made of ultra low expansion glass \cite{leg09}. The resonator has a finesse of about 330\;000 and is housed in a gold plated passive heat shield in an ultra high vacuum system that is actively temperature stabilized. The whole reference cavity setup is mounted on a passive vibration isolation platform. A laser linewidth of 1~Hz was observed by comparison with other laser systems via a fs-laser comb \cite{vog11}.

We use an active fibre noise cancellation system to stabilize the optical path length between the clock laser and the retro-reflection mirror of the optical lattice on which the clock laser beam is overlapped with the lattice. This mirror determines the position of the atoms. By stabilizing the path length we ensure a constant phase of the clock laser field at the position of the atoms. The atoms are interrogated by pulsing the light delivered through the fibre with an acousto-optical modulator (AOM). Thus, the stabilization must be operated in a pulsed mode, too. Special care has been taken to avoid transient phase chirps during switching (Sec.~\ref{sec:others}, details will be presented elsewhere \cite{fal11a}).

\begin{figure}
\centerline{\includegraphics[width=1\columnwidth]{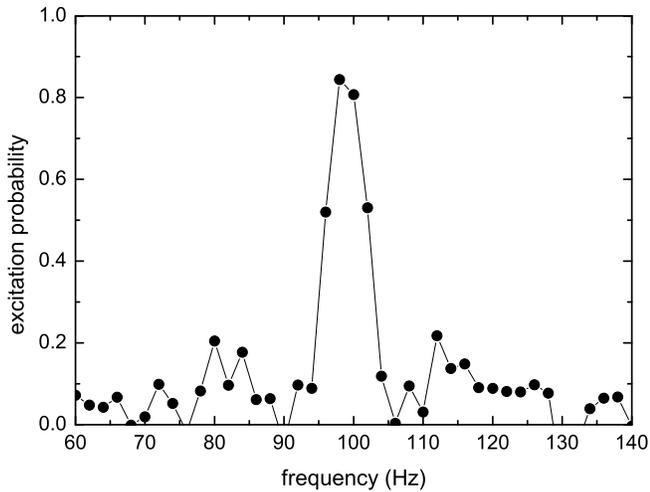}}
\caption{High resolution spectrum of one Zeeman component of $^{87}$Sr excited with a 90~ms spectroscopy pulse. A Fourier limited linewidth of 9~Hz and an excitation probability of about 85\% is achieved. Each point is a single measurement requiring 0.625~s.}
\label{fig:line}
\end{figure}

Having prepared typically $2 \times 10^4$ $^{87}$Sr atoms in a defined $m_F$ level of the $^3$P$_0$ state by the $\pi$ pulse described above, atoms remaining in the $^1$S$_0$ state are pushed away by radiation pressure from a resonant 461~nm laser beam. Then the $\Delta m_F = 0$ transition starting from the prepared extreme $m_F$ level is interrogated in a homogeneous field of 23~$\mu$T with a $\pi$-pulse of typically 90~ms length. We observe Fourier limited spectra with high (de-)excitation probability (Fig.~\ref{fig:line}). To determine the transition probability we, first, detect $^1$S$_0$ atoms by exciting the $^1$S$_0$ -- $^1$P$_1$ transition and detecting the fluorescence with a photo diode. Then the ground state atoms are removed and atoms in the $^3$P$_0$ state are pumped via the $^3$S$_1$ state to the $^1$S$_0$ ground state with high efficiency. Finally, a second fluorescence detection is applied for these atoms. In this way, we can determine the transition probability regardless of shot-to-shot atom number fluctuations.

To lock the clock laser to the unperturbed $^1$S$_0$ -- $^3$P$_0$ transition frequency, the $\Delta m_F = 0$, $m_F = \pm9/2$ lines are each interrogated on the high and low frequency points of maximum line slope i.e. at a detuning of about $\pm 5$~Hz from their respective line centres. The frequency shifts to address the Zeeman components are introduced by the AOM used to pulse the light of the clock laser. The magnetic field causes a splitting of the two lines of about 227~Hz. With the four excitation probabilities the clock laser can be steered to the mean of the two line centres, which is in first order Zeeman-shift-free. The locking is achieved by changing the frequency of the offset AOM between laser and cavity. From the observed transition probabilities the Zeeman splitting is monitored and tracked, too. 

Under the conditions described above a single interrogation cycle requires $\sim$625~ms. We first interrogate one Zeeman component on both sides of its maximum and then switch to the second one. After these four interrogation cycles we correct the determined frequency offset. To minimize locking errors, the drift of the clock laser due to the drift of its reference cavity is zeroed by a continuous frequency sweep on the offset AOM between laser and cavity. The frequency of the direct digital synthesizer (DDS) is altered in small steps (resolution 1.4~$\mu$Hz) at a rate of 100~Hz to obtain a slow quasi-linear ramp to compensate the initially estimated cavity drift rate. We update the sweep rate of the DDS after every interrogation of the four points. Thus the light of the clock laser is at a fixed frequency with a constant offset from the $^{87}$Sr clock transition given by the switching AOM to the experiment. This light is delivered via an actively length stabilized optical fibre to a fs~frequency comb, where the frequency of the clock laser is measured against a H-maser whose frequency is continuously measured by the PTB Cs fountain clock CSF1 \cite{wey01,wey02}.

\begin{figure}
\centerline{\includegraphics[width=1\columnwidth]{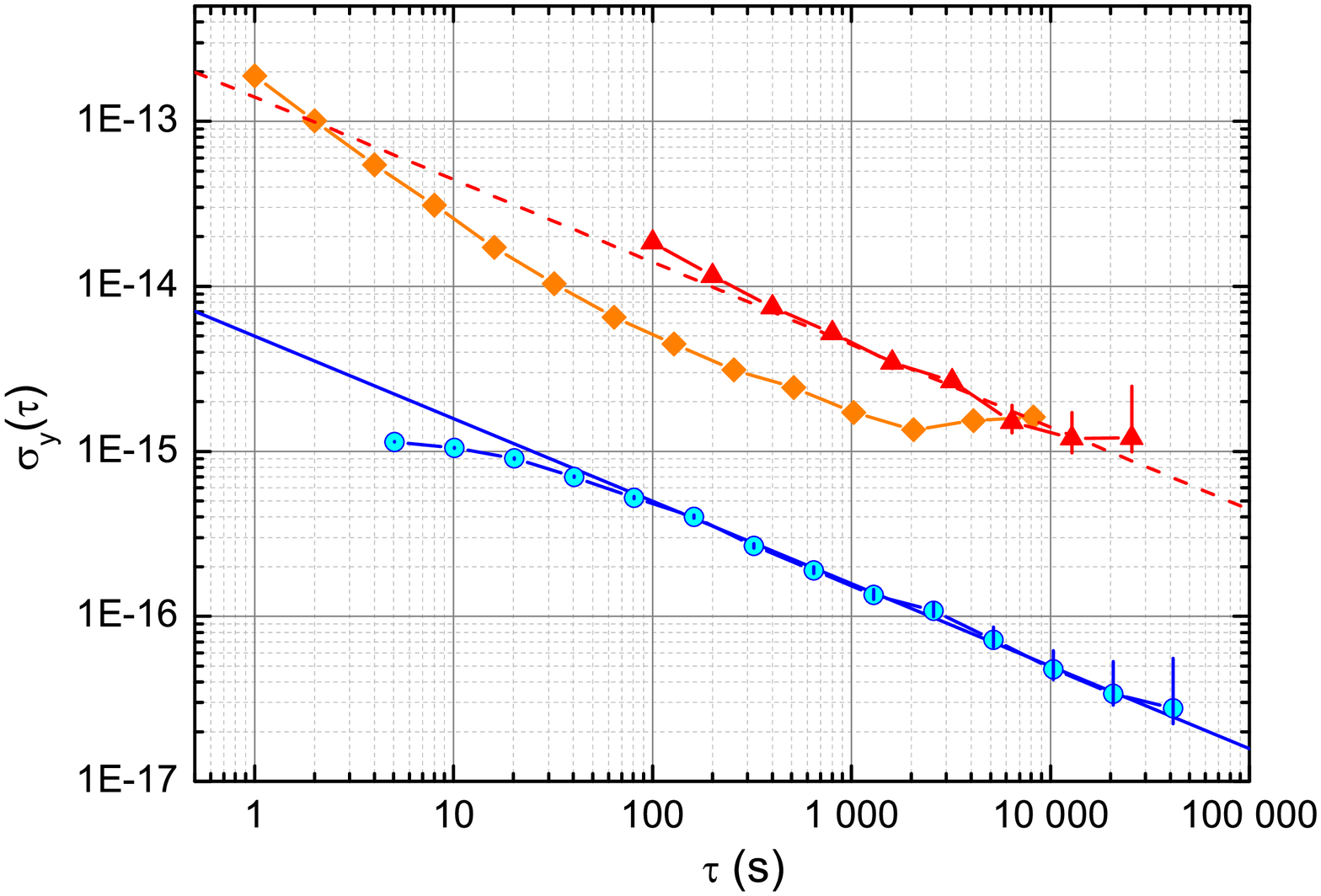}}
\caption{Total fractional Allan deviation of the interleaved stabilization signal (dots, cumulated data set of all days), the frequency measurement against the H-maser (diamonds, one record), and of the cumulated measurement against CSF1 (triangles). The line indicates the stability of the interleaved signal of $5 \times 10^{-15}/\sqrt{\tau/{\rm s}}$, in which common mode effects do not show up. However, the uncertainty budget suggests that no effects occur down to the $10^{-16}$ level that degrade the stability. The dashed line shows the stability of CSF1 of $1.4 \times 10^{-13}/\sqrt{\tau/{\rm s}}$.}
\label{fig:allan}
\end{figure}

For systematic investigations we interleave two stabilization sequences with four interrogations in each case \cite{deg05, deg05a}. With this method we measure systematic frequency changes due to different physical conditions in both sequences. This method has been applied also during the actual frequency measurement with equal sequences. The difference between the two stabilized frequencies provides information on possible locking errors and the quality of the stabilization of the clock laser on the atomic resonance. The observed fractional stability of the difference frequency follows the dependence $5 \times 10^{-15} / \sqrt{\tau/{\rm s}}$ (see Fig.~\ref{fig:allan}), which suggests a possible stability of $2.5 \times 10^{-15} / \sqrt{\tau/{\rm s}}$ if all interrogations are utilized for one laser stabilization. No significant frequency offset between the two interleaved stabilizations was detected.

%
%
%%%%%%%%%%%%%%%%%%%%%%%%%%%%%%%%%%%%%%%%%%%%%%%%%%%%%%%%%%%%%%%%%%%%%%%%%%%%%%%%%%%%%%%%%%%%%%%
%
%	Systematic shifts
%
%%%%%%%%%%%%%%%%%%%%%%%%%%%%%%%%%%%%%%%%%%%%%%%%%%%%%%%%%%%%%%%%%%%%%%%%%%%%%%%%%%%%%%%%%%%%%%%
%
%
\section{Systematic shifts of the $^{87}$Sr transition} \label{sec:shift}

The interleaved stabilization method (see Sec.~\ref{sec:setup}) allows to determine frequency shifts with a fractional statistical uncertainty of $10^{-16}$ after 2\;500~s (Fig.~\ref{fig:allan}). Similar uncertainty contributions have been evaluated in other groups with mostly comparable results \cite{lud08, cam08b, bai08, hon09}. The various corrections and uncertainty contributions are listed in Tab.~\ref{tab:unc} and will be discussed in detail below.
%
%%%%%%%%%%%%%%%%%%%%%%%%%%%%%%%%%%%%%%%%%%%%%%%%%%%%%%%%%%%%%%%%%%%%%%%%%%%%%%%%%%%%%%%%%%%%%%%
%	ac Stark shifts
%%%%%%%%%%%%%%%%%%%%%%%%%%%%%%%%%%%%%%%%%%%%%%%%%%%%%%%%%%%%%%%%%%%%%%%%%%%%%%%%%%%%%%%%%%%%%%%
\subsection{ac Stark shifts} \label{sec:acStark}
Frequency shifts of the clock transition due to ac Stark shifts lead to the most prominent contributions to the corrections and uncertainties listed in Tab.~\ref{tab:unc}. The light shift of the lattice can be classified into first-order shifts with linear dependence on the trap depth, hyperpolarizability contributions \cite{bru06, wes11} (quadratic in trap depth) and contribution due to M1 or E2 transitions \cite{tai08, wes11}. 

\begin{figure}
\centerline{\includegraphics[width=1\columnwidth]{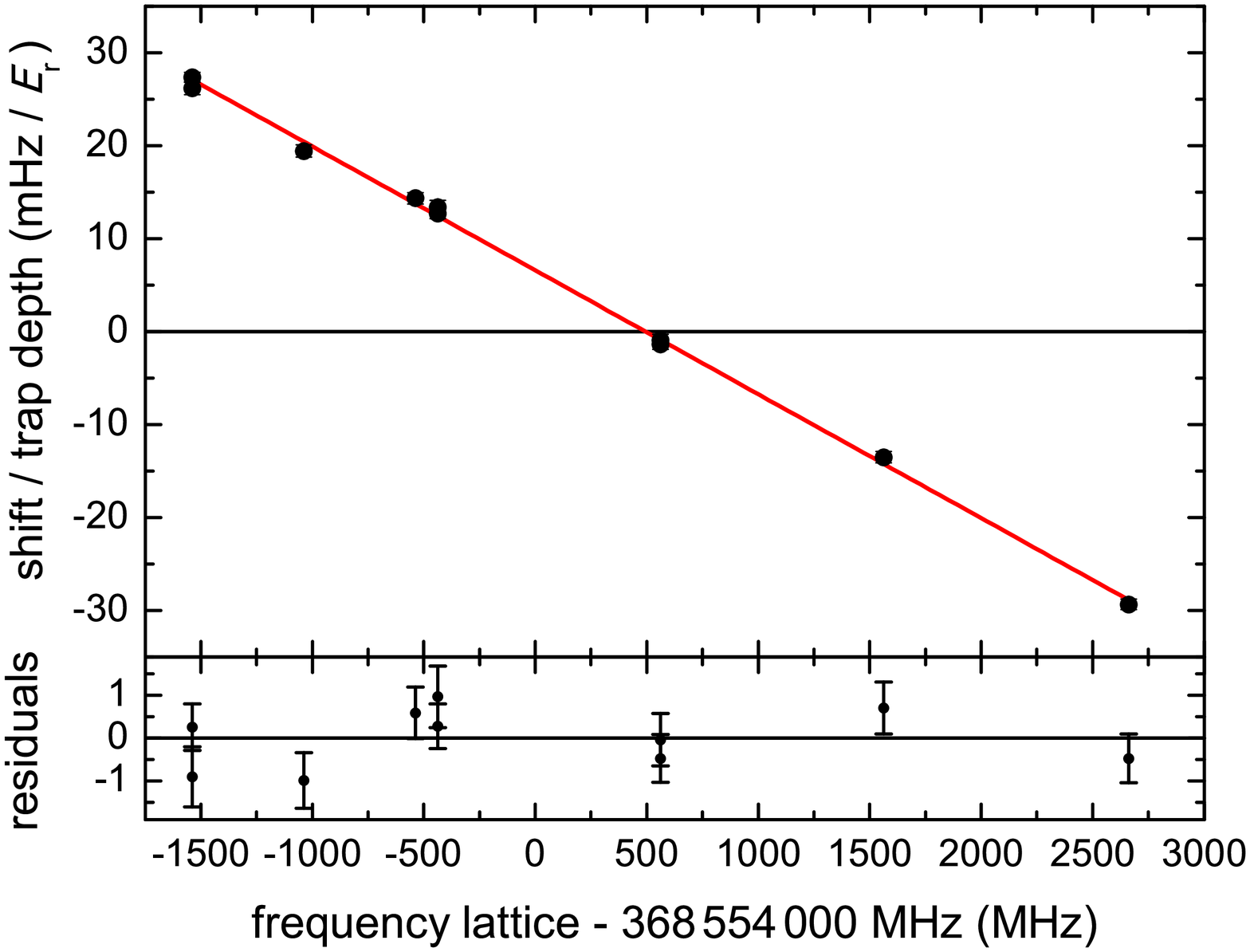}}
\caption{Clock transition shift measured with an interleaved stabilization between a trap depth of  $\sim$130~$E_r$ and $\sim$350~$E_r$ (circles, upper panel). The line is a linear fit to the data to determine the zero crossing for $|m_F| = 9/2$. Lower panel: residuals of the fit with error bars. The zero crossing is found to be at a lattice laser frequency of 368\;554\;502(15)~MHz.}
\label{fig:lat}
\end{figure}

We have determined the (linear) Stark shift cancellation frequency of the lattice by variation of its frequency, using an interleaved stabilization with variation of the lattice depth between $\sim$130~$E_r$ and $\sim$350~$E_r$. The observed frequency shifts are corrected for a small contribution from the hyperpolarizability \cite{wes11}. The shifts are shown as function of the lattice laser frequency in Fig.~\ref{fig:lat}. They have been normalized to the difference of the average lattice trap depths experienced by the atoms. This averaging depends on the lattice geometry (deduced from the vibration frequencies) and atom temperature i.e. the mean particle energy in the potential \cite{bla09}. The temperature is determined from the intensity ratio of red and blue axial sideband. The measured sideband frequencies were in agreement with the ones calculated using the measured power in the lattice and the polarizability at the magic wavelength \cite{por06, por08}. From this agreement we conclude that the lattice has a modulation depth as expected for perfect spatial overlap and that a contribution to the light shift from imperfect interference of the lattice beams can be neglected. 

\begin{table}
\caption{Frequency corrections and their uncertainties}
\label{tab:unc}
\begin{tabular}{lcc}
\hline
Effect	& Correction		& Uncertainty \\
      	&  ($10^{-16}$)	& ($10^{-16}$)\\
\hline
lattice ac Stark (scalar \& tensor)	&  $-0.2$ & 0.5 \\
hyperpolarizability					& 0 & 0.2 \\
E2/M1 Stark shifts					& 0 & 0.2 \\
ac Stark effect probe laser	& 0.2 & 0.2 \\
BBR ac Stark (room temperature)								& 52.5 & 1.3 \\
BBR ac Stark  (oven)					& 0.2 & 0.2 \\
2nd order Zeeman						& 0.28 & 0.03 \\
collisions									& 0.1 & 0.2 \\
gravitation									& 0.1 & 0.1 \\
line pulling								& 0 & 0.4 \\
servo error									& 0 & 0.06 \\
AOM chirp										& 0 & 0.2 \\
efficiency switch AOM				& 0 & 0.02 \\ 
tunneling										& 0 & 0.16 \\
2nd order Doppler						& 0 &  $\ll 10^{-18}$\\
 \hline
\bf{total Sr}								& \bf{53.2} & \bf{1.5} \\
 \hline
fs~comb	and 200~m rf link	  & 0 & 1.5 \\
rf electronics						  & 0 & 4.0 \\
realization of the second   & -- & 7.6 \\
measurement statistics (60\;500~s)							& -- & 5.7 \\
\hline
\bf{total all}							& \bf{53.2} & \bf{10.5} \\
\hline
\end{tabular}
\end{table}

It must be noted that the actual Stark shift cancellation frequency depends on the polarization of the light field, its orientation with respect to the magnetic field axis, and the Zeeman components considered \cite{cam08b, boy07a}. We avoid possible vector light shift contributions due to the linear polarization of the lattice light field and its parallel orientation with the magnetic field axis plus the averaging of the transition frequencies of $m_F = \pm9/2$ components. We find the corresponding cancellation frequency to be at 368\;554\;502(15)~MHz. The value is in agreement with the values given in \cite{lud08,cam08b}. During the frequency measurement the lattice was operated at a 9~MHz lower frequency. We apply a correction of $-2(5) \times 10^{-17}$ based on the mean lattice field seen by the atoms and the slope of the line in Fig.~\ref{fig:lat}. We find a slope of the curve in Fig.~\ref{fig:lat} that is smaller than the slope determined by other groups \cite{pierre10}. This inconsistency has however  otherwise only a minor influence. The uncertainty reflects the inconsistency and the accuracy of the determination of the magic wavelength.

The ac Stark shifts due to the hyperpolarizability and M1/E2 transitions are small \cite{wes11}. We use the trap depth of 125~$E_r$ during the frequency measurement to calculate an upper bound for these shifts. We deduce for both cases an uncertainty of $2 \times 10^{-17}$.

The ac Stark effect of the probe laser is very small and can be calculated with sufficient accuracy. With a clock laser intensity of less than 0.6~mW/cm$^2$ and the coefficient from Ref.~\cite{bai07} of $\delta_{\rm clock} = 13(2)$~Hz/(W/cm$^2$) we find an upper limit of $2(2) \times 10^{-17}$ for the correction. The beams of repump and cooling lasers are blocked by mechanical shutters and the vacuum chamber is shielded from remaining stray laser light. These measures allow neglecting possible ac~Stark shifts by laser beams other than the ones from the clock and lattice laser.

An important correction in the case of Sr frequency standards is due the blackbody radiation (BBR) shift. The shift close to room temperature is according to \cite{por06}
\begin{equation}
 \delta \nu_{\rm BBR} = -2.354(32) \left( \frac{T}{300\: \rm{K}} \right)^4 \: \rm{Hz} \label{eq:bbr}.
\end{equation}
To the uncertainty from theory, the uncertainty of the temperature measurement has to be added. We have continuously monitored the temperature of the vacuum system at ten positions with Pt100 sensors. Typical temperature variations over a day are at each point smaller than 100~mK, however peak-to-peak gradients of 4.2~K are observed over the vacuum system. The tolerance of the Pt100 sensors add an uncertainty of about 0.5~K. We calculate the temperature and its uncertainty assuming that it lies with equal probability an interval of 5.2~K around the best estimate of 296.8~K (rectangular probability distribution, \cite{gum95}) and find a BBR correction of $5.25(13) \times 10^{-15}$.

Further, we consider the influence of radiation coming from the hot furnace. Though it has no direct line-of-sight to the atoms, blackbody radiation could be multiply scattered in the water-cooled Zeeman slower tube, enter the vacuum system and be reflected onto the atoms. In a worst case scenario this can be approximated by a virtual orifice placed on the wall of the vacuum system that has the area of the collimation pinhole shielding the furnace BBR (Sec.~\ref{sec:setup}) and that emits BBR of 793~K homogeneously into the vacuum system.

Using the models described in Ref.~\cite{mid11} this radiation causes additional shifts by the direct line-of-sight between orifices and atoms of 
\begin{equation}
\delta \nu^{\rm (dir)}_{\rm BBR}(793\;{\rm K}) \approx -\frac{\Theta}{4 \pi}\times 146\;{\rm Hz} \label{eq:dir}
\end{equation}
and by the diffusely scattered radiation of
\begin{equation}
\delta \nu^{\rm (diff)}_{\rm BBR}(793\;{\rm K}) \approx -\frac{\Theta}{4 \pi} \frac{1}{a} \times 146\;{\rm Hz} \label{eq:diff}.
\end{equation}
We have introduced $\Theta$ as the solid angle under which the atoms see the virtual orifice and $a$, the absorption coefficient of the walls. Using the diameter of our vacuum system of 335~mm and assuming $a \approx 0.2$ \cite{nam82}, we find $\left| \delta \nu^{\rm (dir)}_{\rm BBR} \right| < 1$~mHz and $\left| \delta \nu^{\rm (diff)}_{\rm BBR} \right| < 7$~mHz. These corrections are on the level of $2\times 10^{-17}$.
%
%%%%%%%%%%%%%%%%%%%%%%%%%%%%%%%%%%%%%%%%%%%%%%%%%%%%%%%%%%%%%%%%%%%%%%%%%%%%%%%%%%%%%%%%%%%%%%%
%	Zeeman
%%%%%%%%%%%%%%%%%%%%%%%%%%%%%%%%%%%%%%%%%%%%%%%%%%%%%%%%%%%%%%%%%%%%%%%%%%%%%%%%%%%%%%%%%%%%%%%
\subsection{Zeeman effect} \label{sec:Zeeman}
As the laser is locked to the average frequency of the $\Delta m_F = 0$ Zeeman transitions from levels $m_F = +9/2$ and $m_F = -9/2$ the linear Zeeman shift is removed for constant fields. Drift in the field give rise to a residual first order Zeeman shift. Great care has been employed that the magnetic field is the same when both Zeeman components are probed. Nevertheless small drifts of the offset field have been observed as temporal variations of the frequency splitting of the two interrogated Zeeman lines. They cause a frequency drift of the two Zeeman lines in opposite direction, which is less than 0.23~mHz per single loading and interrogation cycle. This causes a frequency offset of only 0.23~mHz after the four interrogations because the effect is mostly compensated by the opposite frequency change. We use this value as uncertainty for the residual linear Zeeman shift and have an uncertainty of $5 \times 10^{-19}$, which we neglect.

\begin{figure}
\centerline{\includegraphics[width=1\columnwidth]{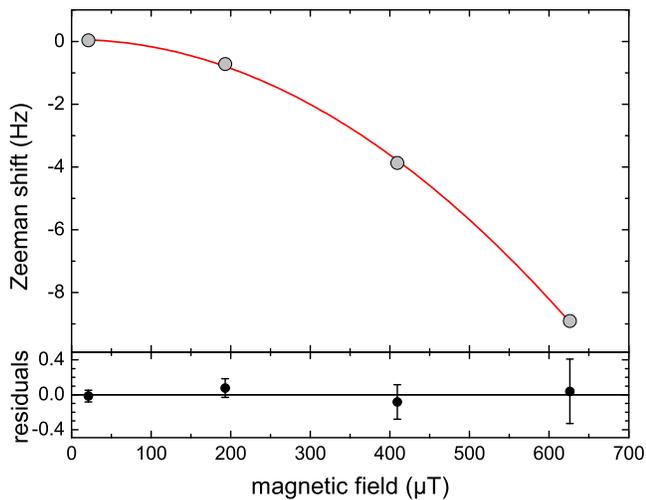}}
\caption{Measurement of the second order Zeeman shift (top, dots) with a parabola fit of the type $aB^2 +c$ and residuals (bottom).}
\label{fig:Zeeman}
\end{figure}

The stabilization sequence provides a value of the Zeeman splitting, which is needed to correct for the second order Zeeman shift. We have varied the homogeneous magnetic field during the spectroscopy pulse between 23~$\mu$T and 630~$\mu$T and have measured the Zeeman shift with the interleaved stabilization against a reference of 23~$\mu$T. The magnetic field was calibrated using the linear Zeeman shift coefficient given in \cite{boy07a}. Figure~\ref{fig:Zeeman} shows the measured shifts in the upper panel together with a fit of the type $aB^2 +c$. In the lower panel the fit residuals are shown. We find for the quadratic coefficient $a = -23.0(3)$~Hz/mT$^2$ which is in good agreement with the value from Ref.~\cite{boy07a}. For the magnetic field of 23(1)~$\mu$T during the frequency measurements we obtain a correction for the second order Zeeman shift of $0.28(3) \times 10^{-16}$.
%
%%%%%%%%%%%%%%%%%%%%%%%%%%%%%%%%%%%%%%%%%%%%%%%%%%%%%%%%%%%%%%%%%%%%%%%%%%%%%%%%%%%%%%%%%%%%%%%
%	Density shift
%%%%%%%%%%%%%%%%%%%%%%%%%%%%%%%%%%%%%%%%%%%%%%%%%%%%%%%%%%%%%%%%%%%%%%%%%%%%%%%%%%%%%%%%%%%%%%%
\subsection{Density shift} \label{sec:density}
\begin{figure}
\centerline{\includegraphics[width=1\columnwidth]{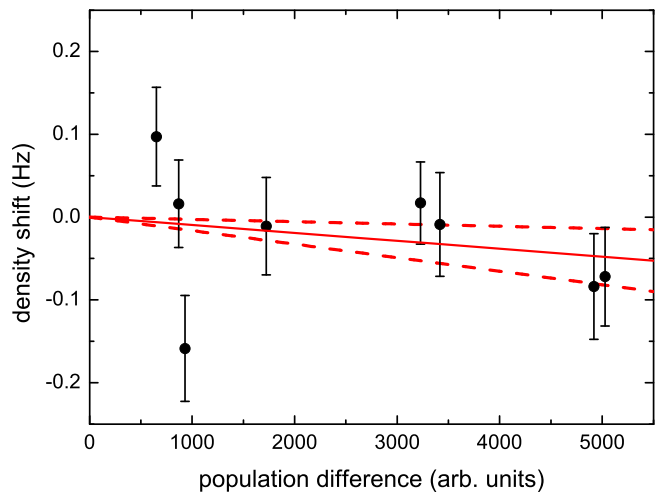}}
\caption{Measurement of the density shift (dots) and linear fit without an offset (line) and its statistical uncertainty (dashed lines). The frequency measurements were performed at a population of about 600 in these arbitrary units, which corresponds to about $10^4$ atoms. The error bars indicate the statistical uncertainty of the interleaved stabilization signal of the corresponding measurement.}
\label{fig:density}
\end{figure}

Observations of a density shift with spin polarized fermions and their theoretical interpretation have triggered intense discussions \cite{cam09,gib09,rey09a,swa11}. A collision shift is observed if different atoms experience different Rabi excitations due to e.g. spatial inhomogeneous excitation fields or motional state dependent Rabi frequencies. Care has been taken to avoid these issues in our experiment by careful alignment and high beam quality. To investigate a possible density shift we vary the atom number in the lattice by a variation of the length of the first MOT phase. We have not observed a significant dependence of the temperature in the lattice on the atom number and therefore assume the density to be proportional to the atom number. In our setup, the determination of absolute atom numbers can only be done with large uncertainties with $^{87}$Sr, but relative atom numbers can be determined reproducibly with high precision. 
We have increased the atom number by nearly a factor of ten compared to the conditions during the frequency measurements and have measured a possible collision shift with respect to a reference density comparable to the one of the frequency measurement. The data are plotted in Fig.~\ref{fig:density}. 
%For the relevant conditions we deduce a correction of $0.1(2) \times 10^{-16}$. 
Theory suggests \cite{gib09} that a density shift may sensitively depend on the excitation parameters which contributes to the scatter of the data.
The uncertainty of the fitted slope shown in Fig.~\ref{fig:density} reflects this effect.
During the frequency measurement the excitation probability was not as well controlled as during this measurement of the collisional shift. Thus, for the correction of the frequency measurement we have increased the uncertainty of the fitted slope by a factor of three and deduce a correction due to collisions of $0.1(2) \times 10^{-16}$. 
%Since we cannot guarantee in this respect perfectly identical experimental conditions for all measurements over a time interval of a few weeks, we have increased the uncertainty of the fitted slope by a factor of three to account for possible variations in the excitation probability.
%Since we cannot guarantee in this respect perfectly identical experimental conditions for the time interval between density %shift measurement and the frequency measurement, we assign an uncertainty of the density shift of twice its value as derived from the fit shown in Fig.~\ref{fig:density}. This is an increase by a factor of three compared to the uncertainty derived from the fit alone.
%
%%%%%%%%%%%%%%%%%%%%%%%%%%%%%%%%%%%%%%%%%%%%%%%%%%%%%%%%%%%%%%%%%%%%%%%%%%%%%%%%%%%%%%%%%%%%%%%
%	others
%%%%%%%%%%%%%%%%%%%%%%%%%%%%%%%%%%%%%%%%%%%%%%%%%%%%%%%%%%%%%%%%%%%%%%%%%%%%%%%%%%%%%%%%%%%%%%%
\subsection{Other corrections and uncertainties} \label{sec:others}
Further corrections have to be applied: The effective position of the atoms in the Cs fountain is 11(10)~cm above the position of the strontium atoms, which leads to a correction of $0.1(1) \times 10^{-16}$ due to the gravitational red shift. The height difference was determined by triangulation between the two buildings that are separated by about 200~m. 

The magnetic field applied during spectroscopy causes a Zeeman splitting of $\sim$25~Hz between neighbouring lines. Line pulling can appear due to atoms in undesired Zeeman levels that are not removed during the preparation sequence. From the achieved signal quality (Fig.~\ref{fig:line}) we can rule out a population fraction in these levels of larger than 5\%, which leads to an uncertainty due to line pulling of $4 \times 10^{-17}$.

The steering of the clock laser to the atomic resonance frequency is compromised by a cavity drift that varies in time. Locking errors occur if the cavity drift rate changes too rapidly. We have observed changes of the cavity drift rate $\ddot{\nu}_{\rm cav}$ of up to $2\times10^{-5}$~Hz/s$^2$. The lock of the laser to the reference line employs two parts. First, a fraction $G = 0.7$ of the measured frequency offset $\nu_{\rm err}$ is applied directly as a correction. Second, the drift rate is altered by $\beta \nu_{\rm err}/\tau_{\rm cycle}$ with $\beta = 0.05$. The time constant for the settling of the drift rate correction is
\begin{equation}
\tau_{\rm drift}=\frac{-\tau_{\rm cycle}}{\ln\left(1-\frac{\beta}{G}\right)} \hspace{2ex} \xrightarrow[\ G\gg \beta\ ]{} \hspace{2ex} \tau_{\rm cycle}\frac{G}{\beta}\ ,
\end{equation}
which amounts to 35~s for our cycle time $\tau_{\rm cycle}=2.5$~s. For a cavity with a constant change of the drift rate of $\ddot{\nu}_{\rm cav}$ the frequency error $\nu_{\rm lock\: err}$ is determined based on the assumption that in steady state the correction of the drift rate has to match the change in the cavity drift that occurs within that lock cycle:
\begin{equation}
\ddot{\nu}_{\rm cav} \cdot \tau_{\rm cycle} = \frac{\beta}{\tau_{\rm cycle}} \; \nu_{\rm lock\: err}\ \ \ \ \ \Rightarrow\ \ \nu_{\rm lock\: err} = \ddot{\nu}_{\rm cav} \frac{{\tau_{\rm cycle}}^2}{\beta}\ .
\end{equation}
For the given experimental conditions this locking error leads to a very small uncertainty of $6 \times 10^{-18}$.

We mentioned in Sec.~\ref{sec:setup} that we have implemented a fibre length stabilization that operates in a pulsed mode. It is used to correct for phase errors due to expansion of the fibre, AOM effects \cite{deg05}, movement of the lattice sites holding the atoms with respect to the clock laser, or other effects that change the interrogation phase of the clock laser at the position of the atoms. However, the servo loop can be activated only at the beginning of the spectroscopy pulse when an error signal is available. This leads to a phase transient before the servo loop settles that causes phase chirps at the position of the atoms. We have measured these chirps with a heterodyne detection and have calculated the resulting frequency offsets using a numerical integration of the time evolution of the density matrix \cite{fal11a}. The remaining uncertainty due to phase errors is $2 \times 10^{-17}$.

The frequency detuning of the clock laser to address the Zeeman components is accomplished by the switching AOM frequency by about 5~Hz. This may result in a systematic variation of the laser intensity and thus systematically different excitation probabilities at the lock points. By measuring the light power transmitted through the fibre delivering the clock laser light to the atoms as a function of AOM frequency we found that this introduces only a marginal uncertainty of $2 \times 10^{-18}$.

Many Sr experiments use vertically oriented optical lattices to avoid tunneling in shallow traps by shifting of the lattice sites by the gravitational potential \cite{lem05}. We have chosen a horizontal orientation to achieve better spatial overlap between MOT and lattice and thus to load a larger number of atoms into the lattice. This requires however a deeper potential to avoid tunneling between the sites and the connected broadening of the band structure.

With the experimental parameters 77\% of the atoms populate the two lowest bands of the lattice. We use the half width of the first excited band of the lattice as upper bound for the effect of tunneling leading to an uncertainty contribution of $1.6 \times 10^{-17}$. 

As last uncertainty contribution we consider the second order Doppler effect of the atoms in the lattice. This effect is however $\ll 10^{-18}$ and completely negligible. 

All corrections discussed in this section add up to $53.2 \times 10^{-16}$ with an uncertainty of $1.5 \times 10^{-16}$ (see Tab.~\ref{tab:unc}).
%
%
%%%%%%%%%%%%%%%%%%%%%%%%%%%%%%%%%%%%%%%%%%%%%%%%%%%%%%%%%%%%%%%%%%%%%%%%%%%%%%%%%%%%%%%%%%%%%%%
%
%	Frequency measurement and results
%
%%%%%%%%%%%%%%%%%%%%%%%%%%%%%%%%%%%%%%%%%%%%%%%%%%%%%%%%%%%%%%%%%%%%%%%%%%%%%%%%%%%%%%%%%%%%%%%
%
%
\section{Frequency measurement and results} \label{sec:frequ}

The frequency of the PTB $^{87}$Sr frequency standard was measured on three days in October 2010. The frequency of the clock laser was measured with a 1.5~$\mu$m Er-doped fs~fibre frequency comb referenced to a 100~MHz rf signal delivered by a semi-rigid cable with low temperature sensitivity from a H-maser. The 5~MHz output of the H-maser was continuously measured by the Cs fountain clock CSF1. The stability (indicated by the total Allan deviation \cite{gre99}) between the Sr frequency standard and the H-maser or fountain clock is shown in Fig.~\ref{fig:allan}.

Both beat frequencies from the Sr clock laser with the fs~comb and of the $f - 2f$ interferometer generating the carrier envelope offset frequency were tracked twice by tracking oscillators. The signals of all tracking oscillators were counted as was the repetition rate of 100~MHz. The double tracking was applied to identify rare cycle slips of phase locked loops and discard them from the data set. Frequency comb measurements connecting microwave and optical frequencies have been demonstrated by several groups at an accuracy level of $10^{-17}$ or below \cite{gro08, mil09a} for averaging times exceeding 1000~s. In the measurements shown here, however, the rf-reference was delivered to the frequency comb via a 200~m semi-rigid cable, which connects the Cs-fountain clock system (in building A) with the frequency comb and Sr-frequency standard (in building B). We thus investigated this 200~m cable with respect to the phase stability of the delivered 100~MHz reference signal; at an averaging time of 10\;000~s, we observe frequency fluctuations (Allan deviation) of about $1.5 \times 10^{-16}$.
 
To give an upper bound for the uncertainty arising from the frequency comb measurement, including the cable connection, we additionally performed simultaneous measurements of an optical frequency (from a cavity locked laser in building B) versus an rf-reference (a H-Maser in building A) with two frequency combs located in the two different buildings. Measurements by the two fs-combs agreed at the level of statistical uncertainty of $1.5 \times 10^{-16}$ for an averaging time of 10\;000~s, which suggests that they were limited in stability by the rf-link discussed above.
During the three days when the Sr-clock transition frequency was measured the air conditioning system in the room of the fountain clock gave rise to periodic 0.2~K peak-to-peak temperature variations with periods in the range of 10\;000~s. An investigation of thereby induced phase shifts of the fountain rf electronics results in an uncertainty contribution of less than $4 \times 10^{-16}$ taking into account the periodicity of the variations and the durations of the single measurements.

\begin{figure}
\centerline{\includegraphics[width=1\columnwidth]{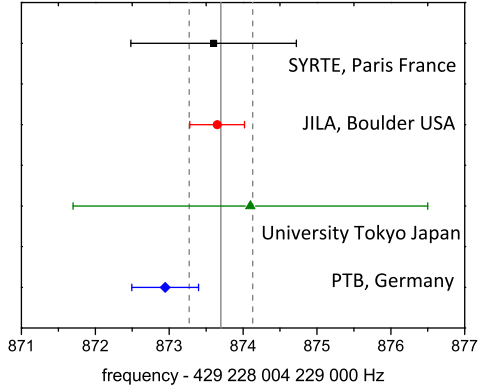}}
\caption{Frequencies of the $^{87}$Sr clock transition measured by different laboratories: Paris (square, \cite{bai08}), Boulder (circle, \cite{cam08b}), Tokyo (triangle, \cite{hon09}), and Braunschweig (diamond). The vertical line gives the recommendation for $^{87}$Sr as secondary representation of the second \cite{cip09} with its uncertainty (dashed lines).}
\label{fig:freq}
\end{figure}

The Cs fountain contributes with a systematical uncertainty of $7.6 \times 10^{-16}$. The fountains instability of $1.4 \times 10^{-13}/\sqrt{\tau/ {\rm s}}$ and the 60\;500~s total record length of the three days measurement lead to a statistical uncertainty of $5.7 \times 10^{-16}$. The measured frequencies of all days are in agreement with each other. The weighted average of the three frequency values is 429\;228\;004\;229\;872.9(5)~Hz. It is in good agreement with previous measurements \cite{lud08, cam08b, bai08, hon09} and the frequency value recommended for the secondary representation of the second by $^{87}$Sr \cite{cip09} (see Fig.~\ref{fig:freq}).
%
%
%%%%%%%%%%%%%%%%%%%%%%%%%%%%%%%%%%%%%%%%%%%%%%%%%%%%%%%%%%%%%%%%%%%%%%%%%%%%%%%%%%%%%%%%%%%%%%%
%
%	Conclusion
%
%%%%%%%%%%%%%%%%%%%%%%%%%%%%%%%%%%%%%%%%%%%%%%%%%%%%%%%%%%%%%%%%%%%%%%%%%%%%%%%%%%%%%%%%%%%%%%%
%
%
\section{Conclusion} \label{sec:con}
We report on a frequency measurement of the $^1$S$_0$ -- $^3$P$_0$ clock transition in $^{87}$Sr with a fractional uncertainty of $1 \times 10^{-15}$ limited by the comparison with the Cs atomic clock. The measurement is in good agreement with previous results achieved in three other laboratories within the combined uncertainties. The estimated standard uncertainty of our $^{87}$Sr standard by itself was estimated to be below $2 \times 10^{-16}$ which competes well with the uncertainty of the best primary Cs clocks for the realization of the {SI} units $\mbox{s}$ and $\mbox{Hz}$ and, hence, supports the choice of this standard as secondary representation of the second. The reported uncertainty represents the status of the clock during the last frequency measurement and could be further improved. 

The presently largest contributions to the uncertainty, results from the blackbody ac Stark effect, the lattice ac Stark effect, and the line pulling effect with $1.3 \times 10^{-16}$, $0.5 \times 10^{-16}$, and $0.4 \times 10^{-16}$, respectively. Investigations are under way to determine the influence of the blackbody radiation shift \cite{mid11} with the prospect to reduce the associated correction and uncertainty significantly. Tuning the frequency of the lattice laser closer to the magic wavelength during clock operation than was possible during the present measurement could reduce the biggest contribution to the lattice ac Stark effect in Tab.~\ref{tab:unc} below $0.1 \times 10^{-16}$. The uncertainty from the line pulling effect due to atoms in undesired Zeeman levels (mainly the $m_F = \pm 7/2$ levels) can be reduced by a more detailed investigation of its influence in the actual preparation conditions by employing different magnetic fields with the associated different influence on the line pulling effect. Similarly, there is no principal limitation that should prevent to reduce the other contributions to below $10^{-17}$ thereby leading to a combined fractional uncertainty around $10^{-17}$. With several $^{87}$Sr clocks being investigated in different institutes it is expected that an in-depth evaluation of this type of optical clock on the $10^{-17}$ level will be achieved soon. Having different optical clocks based on neutral atoms and on single ions at hand that have an order of magnitude higher accuracy than the best primary Cs clocks of today, the next steps towards a re-definition of the second as the base unit of the SI could be made.

\begin{acknowledgments}
The support by the Centre of Quantum Engineering and Space-Time Research (QUEST), funding from the European Community's ERA-NET-Plus Programme (Grant No. 217257), and by the ESA and DLR in the project Space Optical Clocks is gratefully acknowledged. We also thank M.~Merimaa, L.~Hannemann, Anshuman Vinit, and Wang Qiang for their contributions to the experimental setup. We also thank M.~Misera and A.~Koczwara for expert electronics support.
\end{acknowledgments}

%
%\bibliography{aipsamp}% Produces the bibliography via BibTeX.
%\bibliographystyle{prsty}
%%\bibliographystyle{nar}
%\bibliography{O:/4-3/4-3-Alle/Papers/TeXBib/TeXBi431}

\end{document}